\documentclass{ws-ijmpd}

\usepackage{times}
\usepackage{graphicx,bm,amssymb}
\usepackage{epsfig}
\usepackage{natbib}

\newcommand{\Zsun}{\mbox{$Z_\odot$}}
\newcommand{\msun}{\mbox{$M_\odot$}}
\newcommand{\rsun}{\mbox{$R_\odot$}}
\def\be{\begin{eqnarray}}
\def\ee{\end{eqnarray}}
\def\lsim{\mathrel{\rlap{\lower3pt\hbox{\hskip1pt$\sim$}}
     \raise1pt\hbox{$<$}}} 
\def\gsim{\mathrel{\rlap{\lower3pt\hbox{\hskip1pt$\sim$}}
     \raise1pt\hbox{$>$}}} 

\def\apj{ApJ}
\def\apjl{ApJ}
\def\apss{Ap\&SS}
\def\aap{A\&A}
\def\mnras{MNRAS}
\def\na{New A}
\def\nat{Nature}
\def\bain{Bull.~Astron.~Inst.~Netherlands}
\def\physrep{Phys.~Rep.}

\begin{document}

\markboth{Enrique Moreno M\'endez}
{From GRBs/HNe to BH Binaries}

\title{From Gamma-Ray Bursts/Hypernovae To Black-Hole Binaries}


\author{Enrique Moreno M\'endez}

\address{Argelander-Institut f\"ur Astronomie, Bonn University,
\\Auf dem H\"ugel 71, 53121 Bonn, Germany.
\\Instituto de Astronom\'ia, Universidad Nacional Aut\'onoma de M\'exico,
\\A.P. 70-264, 04510 M\'exico D.F., M\'exico.\\
enriquemm@astro.unam.mx
}

\maketitle

\begin{history}
\received{24 01 2012}
\revised{Day Month Year}
\comby{Managing Editor}
\end{history}

\begin{abstract}
In this work I summarize a model of binary stellar evolution involving Case C mass transfer followed by a common envelope that strips away the hydrogen from the core of the primary star at the cost of shrinking the orbital separation and then, through tidal interaction, spins it up.
This model is then used to produce the possible progenitors of long gamma-ray burst / hypernova (GRB/HN) explosions.
As the core collapses with the newly supplied angular momentum it produces a Kerr black hole surrounded by an accretion disk.
Energy is extracted from the rotation of the black hole (BH) through the Blandford-Znajek (BZ) mechanism to power both, the long gamma-ray burst and the accompanying  hypernova (supernova type Ic broad line).
If the binary survives the asymmetric mass loss its remnant is a black-hole binary that may eventually be observed as a soft X-ray transient (SXT) when the companion evolves and starts to transfer mass back to the black hole.
A comparison with a sample of black-hole binaries where the masses and orbital periods are well constrained is performed.
\end{abstract}


\section{Introduction\label{sec-Intro}}

The Collapsar model \citep{1993ApJ...405..273W,1999ApJ...524..262M} predicts that long-duration, soft spectrum, gamma-ray bursts (GRBs) would be accompanied by a supernova (SN) explosion.  This was first confirmed by the spectroscopic association of GRB 980425 with SN 1998bw and later followed by another four events \citep[GRB 030329/SN 2003dh, GRB 031203/SN 2003lw, GRB 060218/SN 2006aj and GRB 100316D/SN 2010bh; see ref.~][for a recent review]{2011arXiv1104.2274H}.
There is less conclusive, but substantial evidence of GRB/SN association in another two-dozen events.
Nonetheless, most gamma-ray bursts have no observed supernova counterpart, most likely because the greater fraction of them occur at large red shifts and thus make a supernova detection unlikely.
Most supernovae associated with gamma-ray bursts are type Ic (broad line), i.e., they show no hydrogen nor helium lines, and large velocities are inferred for the ejected material ($V_{ejecta} \gtrsim 3 \times 10^4$ km/s).
As pointed out by \citet{1998ApJ...494L..45P}, long gamma-ray bursts are also associated with star forming regions, thus strengthening the relation with the collapse of massive stars given that such stars have lifetime scales that span only a few million years and are, thus, incapable of being displaced  long distances from their birthplaces before they collapse into a black hole.

In the same paper, \citet{1998ApJ...494L..45P} suggests that the central engine of these long-soft gamma-ray bursts is powered up by a compact object with large angular momentum and magnetic fields through the Blandford-Znajek (BZ) mechanism \citep{1977MNRAS.179..433B}.
Almost in parallel, \citet{1998ApJ...506..780B,1999ApJ...517..318B} (and collaborators) worked on the formation of binaries with black holes (BHs), Blandford-Znajek mechanism \citep{2000PhR...325...83L} and gamma-ray bursts \citet{2002ApJ...575..996L}.

In this work a summary is provided of some of the most relevant results resulting from the idea of using the formation of a black hole in a close binary as the source of GRB/HN explosions.
In the next section a brief explanation of some key ideas of the Collapsar model is given.
This is followed by a description of the stellar binary model leading to Case C mass transfer and the spin up of massive stars in the third section.
Next, some relevant results from the Blandford-Znajek model are covered in the fourth section.
This leads to a description of the best known black-hole binaries in our Galaxy and nearby ones in the fifth section.
To finish the description of black-hole binaries, in the sixth section we discuss some massive systems some of which may harbor future gamma-ray bursts.
The conclusions are shown in the last (seventh) section of this work.

\section{Collapsar Model\label{sec-Collaps}}

Massive single stars ($M_{ZAMS}\gtrsim 8 \msun$, where $M_{ZAMS}$ stands for the zero-age-main-sequence mass) evolve while burning through their nuclear-fuel resources and eventually explode as supernovae leaving behind a compact object.
If the ZAMS mass of the star is somewhere below the range between the 18 to 25 $\msun$ it is likely that the remnant will be a neutron star (NS).
But if the ZAMS mass is above this limit, the remnant will likely exceed the maximum mass for a stable neutron star (which is likely between 3 and $5 \msun$ and depends on the equation of state of matter at density values above the nuclear one) and will further collapse and form a black hole.
These mass limits are somewhat more vague if the stars evolve in binaries given that mass transfer \citep[see e.g.][]{1995A&A...297..483B,2007ApJ...671L..41B} and tides \citep[e.g.][]{2009A&A...497..243D} may influence the evolution of the star.

The Collapsar model considers a massive star that collapses and fails to launch a supernova, thus forming a black hole a few seconds after the initial collapse.
Material in the core of the star as well as along the rotation axis of the star, having little to no angular momentum, collapses directly, clearing a path along this axis.
Most of the material outside the central core of the star possesses enough angular momentum to prevent it from directly falling in and, thus, forms an accretion disk.

Before the collapse, the core of the star, where the silicon ashes burn into iron, has a temperature of  $T_c \sim 4 \times 10^9$ K, and thus all the material is completely ionized.
This results in a magnetic field which is {\it frozen} into the material.
As the collapse develops the magnetic flux as well as the available angular momentum are conserved.
Thus, during the contraction phase, the proto-neutron star spins more and more rapidly and its magnetic field increases.

Eventually, the proto-neutron star becomes too massive to prevent the formation of a black hole and it collapses into one.
The interaction between the rapidly spinning black hole at the center of the star and the accretion disk is expected to produce two jets that will finish removing the remaining material along the rotation axis of the star.  This will be observed as a gamma-ray burst.
The outer layers of the accretion disk will be heated up until they either are expelled via a strong wind or they explode.
These will be observed as an extremely luminous and rapidly expanding supernova, often termed {\it hypernova} (HN).

\section{Binary Evolution\label{sec-Binary}}

As it has been shown by, e.g., \citet[][]{2005ApJ...626..350H} and \citet{2006A&A...460..199Y}, a star which is born with
more-than-enough angular momentum in its core to produce a Collapsar, may lose most of this angular momentum by the time it is finished burning hydrogen (i.e., at the end of the main sequence).
Even if this were not the case and enough angular momentum was still left in the star, during the helium-burning stage the envelope of a massive star expands from a few to several-hundred solar radii carrying out most of the angular momentum of the star.
Even though the helium core contracts and spins up, viscosity and/or magnetic fields slow it down (a viscous or an Alfv\'en timescale are usually much shorter than the nuclear timescale) until solid-body rotation is restored, thus draining angular momentum from the stellar core.
During the helium-burning stage and later ones most of the envelope is lost due to strong stellar winds carrying away most of the remaining angular momentum left in the star.
Such estimates are made with the assumption that the Spruit-Tayler dynamo \citep{2002A&A...381..923S} is at work inside the star and, thus, the angular momentum is transported due to magnetic torques; this may not be the complete picture, nonetheless the predictions for the rotation of pulsars and white dwarves seem to be consistent with their corresponding observed populations.

If most massive stars are drained out of their original angular momentum during their evolution then it is clear that Collapsars (gamma-ray burst progenitors) may not evolve through this channel given that, both, a strong magnetic field and a very large amount of angular momentum in the collapsing core are needed, and they seem to be exclusive of each other.

At least two different solutions have been proposed around this problem.
One of them is to have rapidly-rotating (above $\sim50 \%$ Keplerian rotation) massive stars which may evolve chemically homogeneously and, thus, avoid the giant phase.
By doing this they do not carry most of their angular momentum into an extended envelope (of up to hundreds of solar radii) which is later lost to winds. Instead, the winds remove angular momentum from a much more compact surface (a few tens of solar radii), allowing the stars to preserve a substantial fraction of their original angular momentum.
Such scenario may occur for single stars \citep[][]{2005ApJ...626..350H,2006A&A...460..199Y}, or tidally-locked binaries \citep{2009A&A...497..243D}.
The other solution is the one we have put forward in \citet{2002ApJ...575..996L,2007ApJ...671L..41B,2008ApJ...685.1063B,2008ApJ...689L...9M,2011ApJ...727...29M,2011MNRAS.413..183M} and will address in the remainder of this work.
It involves a binary in an initially wide orbit ($a_{orb} \gtrsim 1,500 \rsun$, where $a_{orb}$ stands for the orbital separation) with a massive star ($M_{ZAMS} \gtrsim 20 \msun$) and a non-massive companion star ($0.5$ to $5 \msun$).
The massive star evolves as a single star through its main sequence and helium-core burning stages, expanding from a few solar radii to $\sim 850 \rsun$ during the red giant stage.
It is until the massive star begins helium-shell burning, when it further expands to $\sim 1,000 \rsun$, that it fills its Roche lobe and starts an unstable mass transfer onto the companion star.
Mass transfer at or after helium-shell burning is known in the literature as Case C mass transfer.
As a consequence the orbit decays rapidly and a common envelope ensues.
The orbital separation keeps decreasing at the expense of removing the hydrogen envelope of the primary.
The common envelope ends when the envelope is fully removed, leaving the massive helium core exposed.
At this stage the orbit has been reduced to merely a few solar radii.
It is worth noting here that the situation is already better than that of a single star because by removing all the hydrogen envelope the gamma-ray burst will not have to struggle its way out of it risking being completely quenched.
Furthermore, it naturally explains the lack of hydrogen lines in the spectra of hypernovae-related to gamma-ray bursts (i.e., they are not type II).

In this new configuration, the exposed core is close to filling its Roche lobe and thus the ratio between the stellar radius, $R$, and the orbital separation is not far from unity.
From \citet{1975A&A....41..329Z,1977A&A....57..383Z} we know that the tidal synchronization timescale depends on $(R/a_{orb})^n$ where $n$ is a high power (either 6 or $17/2$ depending on whether the equilibrium tide or the dynamical tide, respectively, are most relevant).
Thus tidal synchronization allows for a quick spin up of the massive star \citep[from $\lesssim 10^3$ to $10^4$ years; see, e.g., the calculation by][]{2007Ap&SS.311..177V}.
At this stage, it is important for the massive star to have a substantial magnetic field ($\gtrsim 10^3$ G) in order to have rapid transfer of angular momentum into the core of the star.

Later in the evolution of the star, an extremely large magnetic field may also steal away part of the angular momentum of the iron core, as estimated by \citet{2011arXiv1110.3842W}.
As we mentioned earlier, the exact transport mechanism for the angular momentum is still not fully understood.
One can estimate values for the internal magnetic field that provide Alfv\'en timescales appropriate to transport inwards angular momentum from the tidal interaction, and where little is lost outwards later, during the shorter lifetimes of the post-carbon-burning stages.

\begin{figure}
\centering
\begin{tabular}{ccc}
\includegraphics[height=55mm,width=0.48\textwidth]{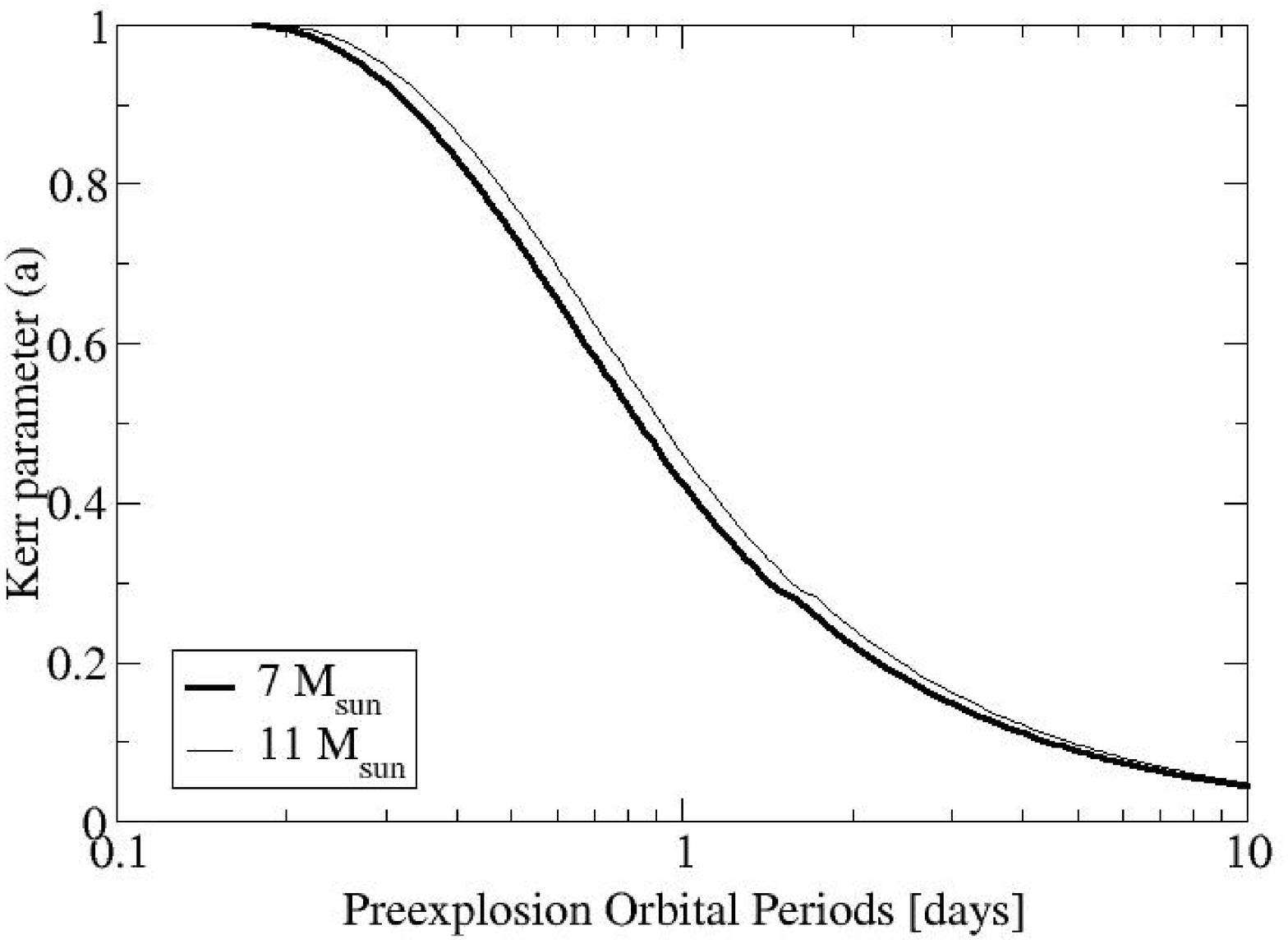}
& \hspace{2mm} &
\includegraphics[height=55mm,width=0.45\textwidth]{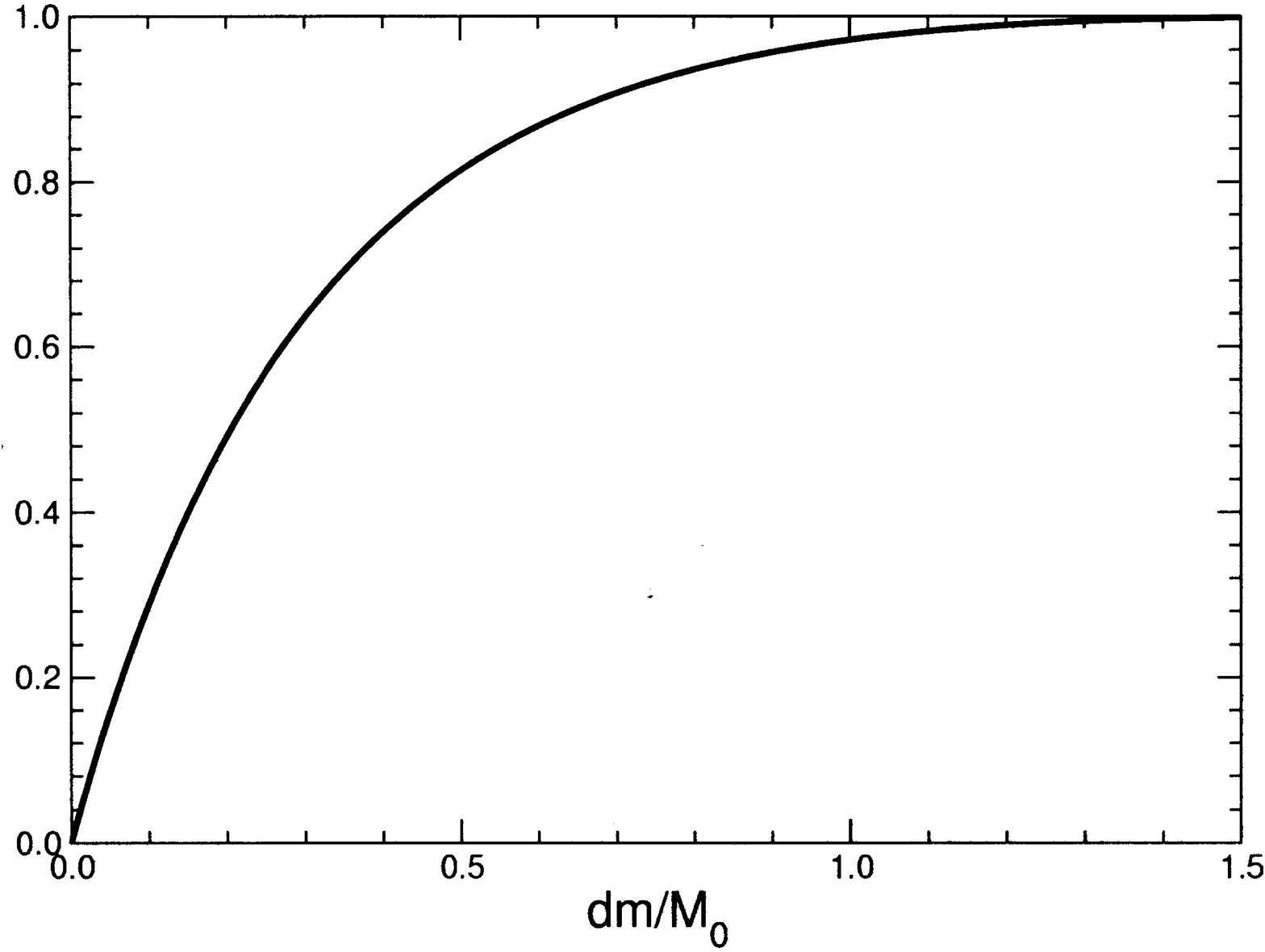}
\end{tabular}
\caption{\small{(Left) The Kerr parameter $a_\star$ of the black hole resulting from the collapse of a He star rotating synchronously with the orbit, as a function of orbital period \cite{2002ApJ...575..996L}. The result depends very little on the mass of the He star, or on whether we use a simple polytrope or a more sophisticated model.
(Right) Spinning up black holes. $a_\star$ is given in units of [$GM/c^2$] and $\delta m$ is the total rest mass of the accreted material. Note that $M_0$ is the mass of the non-rotating initial black hole; we assumed that the last stable orbit corresponds to the marginally stable radius \cite{2000NewA....5..191B}.}} \label{fig:1}
\end{figure}

If the star preserves this amount of angular momentum until it collapses, the left plot of Fig.~\ref{fig:1} \citep[from][]{2002ApJ...575..996L} shows the expected spin or Kerr parameter ($a_\star \equiv Jc/GM^2$ where $J$ is the angular momentum, and $M$ is the mass of the collapsed object) of the rotating black hole as a function of the orbital period of the pre-collapse binary.

\section{Energetics: Blandford-Znajek\label{sec-BZ}}

During the formation of a stellar-mass black hole, the angular momentum as well as the magnetic flux that threads the core of the star are conserved.
If a substantial Kerr parameter (maybe $a_\star \gtrsim 0.25$) is achieved, an accretion disk forms early on and a large magnetic field is produced ($B \sim 10^{15}$ G) the conditions are ripe for a central engine of the Blandford-Znajek \citep{1977MNRAS.179..433B} type to power a GRB/HN explosion.

A spinning black hole of mass $M_{BH}$ and Kerr parameter, $a_\star$, contains a large amount of rotational energy which can be quantified by using the following formulas:
\be
E_{rot} = f(a_\star) M_{BH} c^2, \, \, \, \, \, \,{\rm where} \, \, \, \, \, \,f(a_\star) = 1-\sqrt{\frac{1}{2}\left(1 + \sqrt{1 - a_\star^2}\right)}.
\ee
Here $f(a_\star)$ is a function of $a_\star$ which varies between 0 (for $a_\star = 0$, i.e., a Schwarzschild black hole) and 0.29 (for $a_\star = 1$, a maximally spinning or Kerr black hole).

This rotational energy can be tapped out of the black hole through interaction with the accretion disk via the strong magnetic field that permeates both.

Essentially, the plasma that conforms the accretion disk cannot freeze in the magnetic field that threads it as the later is dragged by the rapidly-rotating black hole, thus producing a Poynting flux along the rotational axis and heating up the plasma in the accretion disk.

At maximum impedance, when the spin frequency of the magnetic field ($\Omega_F$) threading the material at the ISCO (the innermost stable circular orbit) and further out is half of the spin frequency of the black hole ($\Omega_{BH}$), or $\epsilon_\Omega \equiv \Omega_F / \Omega_{BH}$ = 0.5, one obtains the maximum energy-extraction efficiency.  An analytical expression that provides the energy  extracted by the Blandford-Znajek mechanism is given by:
\be
E_{BZ} = 1.8 \times 10^{54} \epsilon_\Omega f(a_\star) \frac{M_{BH}}{\msun} {\rm erg}.
\ee

As we will show in the following sections, this result can be directly applied to several black-hole binaries where the masses and orbital periods are well constrained to obtain the available energies in their black holes.
The more reliable estimates for Kerr parameter and available energy will come from those systems where modeling the evolution of the binary back to the pre-collapse situation can be better constrained because, as we have mentioned above, the pre-collapse orbital period is what determines the Kerr parameter of the black hole.

\section{Black-Hole Binaries\label{sec-BHBs}}

\begin{sidewaystable}[ph]
\vspace{12cm}
\tbl{Parameters at the time of formation of black hole and at present time.  Subindex $2$ stands for values at the time black hole is formed, whereas subindex $now$ stands for recently measured values.  The AML (angular-momentum loss) binaries lose energy by GWs, shortening the orbital period whereas the Nu (Nuclear evolved) binaries will experience mass loss from the donor star to the higher mass black hole and, therefore, move to longer orbital periods. \newline 
REFERENCES:  Most observational data from:  \citet{2002ApJ...575..996L} (and references therein), except for XTEJ1859$+$226 from:  \citet{2011MNRAS.413L..15C}, A0620$-$003 from:  \citet{2010ApJ...718L.122G} (and references therein), XTE J1550$-$564 from:  \citet{2011MNRAS.tmp.1036S} (and references therein), Cyg X$-$1 from: \citet{2011ApJ...742...83R,2011ApJ...742...84O,2011ApJ...742...85G}, LMC X$-$1 from:  \citet{2009ApJ...697..573O,2009ApJ...701.1076G}, LMC X$-$3 from:  \citep{2003IAUS..212..365O,2006ApJ...647..525D} (and references therein) and M33 X$-$7 from:  \citet{2006ApJ...646..420P,2007Natur.449..872O,2008ApJ...679L..37L}.}
{\begin{tabular}{|c|c|c|c|c|c|c|c|c|}
\hline
      Name     & $M_{BH,2}$ & $M_{d,2}$ &  $M_{BH,now}$ & $M_{d,now}$ &     Model     &    Measured   & $P_{b,now}$ &   $E_{\rm BZ}$   \\
               & [$\msun$]  & [$\msun$] &   [$\msun$]   &  [$\msun$]  & $a_{\star,2}$ & $a_{\star}$ &      [days]     & [$10^{51}$ ergs]   \\
\hline
\hline
\multicolumn{9}{|c|}{AML: with main sequence companion} \\
\hline
J1118$+$480    &   $\sim5$  &   $<1$    &    $6.0-7.7$   &   $0.09-0.5$ &    $0.8$   &       -       &  $0.169930(4)$ &   $\sim 430$      \\
Vel 93         &   $\sim5$  &   $<1$    &    $3.64-4.74$ &  $0.50-0.65$ &    $0.8$   &       -       &     $0.2852$   &   $\sim 430$      \\
GRO J0422$+$32 &    $6-7$   &   $<1$    &    $3.4-14.0$  &  $0.10-0.97$ &    $0.8$   &       -       &   $0.2127(7)$  & $500\sim 600$     \\
XTE J1859$+$226&    $6-7$   &   $<1$    &      $>5.42$   &              &    $0.8$   &       -       &    $0.274(2)$  & $500\sim 600$     \\ 
GS1124$-$683   &    $6-7$   &   $<1$    &     $6.95(6)$  &  $0.56-0.90$ &    $0.8$   &       -       &     $0.4326$   & $500\sim 600$     \\
H1705$-$250    &    $6-7$   &   $<1$    &     $5.2-8.6$  &    $0.3-0.6$ &    $0.8$   &       -       &     $0.5213$   & $500\sim 600$     \\
A0620$-$003    &  $\sim10$  &   $<1$    &  $6.61\pm0.25$ &$0.40\pm0.045$&    $0.6$   & $0.12\pm0.19$ &     $0.3230$   &   $\sim 440$      \\ 
GS2000$+$251   &  $\sim10$  &   $<1$    &    $6.04-13.9$ &  $0.26-0.59$ &    $0.6$   &       -       &     $0.3441$   &   $\sim 440$      \\
\hline
\hline
\multicolumn{9}{|c|}{Nu: with evolved companion} \\
\hline
GRO J1655$-$40 &   $\sim5$  &  $1-2$    &    $5.1-5.7$   &   $1.1-1.8$  &    $0.8$   &  $0.65-0.75$  &   $2.6127(8)$  &   $\sim 430$      \\
4U 1543$-$47   &   $\sim5$  &  $1-2$    &    $2.0-9.7$   &   $1.3-2.6$  &    $0.8$   &  $0.75-0.85$  &     $1.1164$   &   $\sim 430$      \\
XTE J1550$-$564&  $\sim10$  &  $1-2$    &  $9.68-11.58$  & $0.96-1.64$  &    $0.5$   & $0.49\pm0.13$ &    $1.552(10)$ &   $\sim 300$      \\ 
GS 2023$+$338  &  $\sim10$  &  $1-2$    &   $10.3-14.2$  & $0.57-0.92$  &    $0.5$   &       -       &     $6.4714$   &   $\sim 300$      \\
XTE J1819$-$254&    $6-7$   &  $\sim10$ &  $8.73-11.69$  & $5.50-8.13$  &    $0.2$   &       -       &      $2.817$   &  $10\sim 12$      \\
GRS 1915$+$105 &    $6-7$   &  $\sim10$ &     $14(4)$    &    $1.2(2)$  &    $0.2$   &    $>0.98$    &    $33.5(15)$  &  $10\sim 12$      \\
Cyg X$-$1      &    $6-7$   &$\gtrsim30$& $14.81\pm0.98$ & $19.2\pm1.9$ &    $0.15$  &    $>0.97$    &     $5.5996$   &    $5\sim6$       \\ 
\hline
\hline
\multicolumn{9}{|c|}{Extragalactic} \\
\hline
LMC X$-$1      &   $\sim40$ &  $\sim35$ &  $8.96-11.64$ & $30.62\pm3.22$& $\sim0.05$ &  $0.81-0.94$  &   $3.90917(5)$ &     $<2$          \\ 
LMC X$-$3      &     $7$    &    $4$    &     $5-11$    &    $6\pm2$    &    $0.43$  &   $\sim0.3$   &   $1.70479(4)$ &    $\sim155$      \\ 
M33 X$-$7      &   $\sim90$ &  $\sim80$ & $14.20- 17.10$&  $70.0\pm6.9$ & $\sim0.05$ &  $0.72-0.82$  & $3.453014(20)$ &      $3-11$       \\ 
\hline
\end{tabular}
\label{tab-results}}
\end{sidewaystable}

As early as \citet{2006astro.ph.12461M} but more in form in \citet{2007ApJ...671L..41B} and \citet{2011ApJ...727...29M}, we studied seven Galactic sources with nuclear-evolved (i.e., beyond main sequence) companion stars (in all cases $M_{comp} \gtrsim 1 \msun$), eight Galactic sources with main sequence companion (all of them with $M_{comp} < 1 \msun$).
We further studied three extragalactic sources \citep{2008ApJ...685.1063B,2008ApJ...689L...9M,2011MNRAS.413..183M} with massive companions.

We reconstructed the configuration at the pre-collapse time from the present-day configuration in order to obtain the Kerr parameters present at black hole formation.
This was done with the idea of obtaining the energies available to the central engine which may have produced a GRB/HN-like event.

It is important to notice that this reconstruction can be tricky in more ways than one.
In systems like the eight Galactic black-hole binaries with main-sequence companions, it is hard to know how much the orbit may have decayed (due to gravitational waves, magnetic torques, etc.) since the formation of the black hole because such low mass companions live for several giga-years and thus the black-hole binary may be that old.
If such is the case, we can expect our results to be upper limits to the Kerr parameters and available energies of the black-hole binaries.

In the case of systems like Nova Sco (GRO J1655$-$40), Il Lupi (4U 1543$-$47), Cyg V404 (GS 2023$+$338), XTE J1550$-$564, Sag V4641 (XTE J1819$-$254),
GRS 1915$+$105 or LMC X$-$3 the story may be a bit more intricate but easier to reconstruct.
In these systems considerable mass transfer may have taken place after the black hole was formed, altering the original $a_\star$.
Nonetheless, the plot on the right-hand side of Fig.~\ref{fig:1} provides help in obtaining the corrected value of the kerr parameter and with some simplifications (e.g., conservative mass transfer) one can estimate the pre-collapse orbital period of the binary.

In general, we find that the pre-collapse orbital period of the binary, $P_b$ (in days), does not depend strongly on the mass of the primary but it does depend strongly on the mass of the companion:
\be
P_b \propto \left(\frac{M_{primary}}{\msun}\right)^{0.83}
\left(\frac{m_{comp}}{\msun}\right)^{3/2}.
\ee
Thus, in general, a smaller companion results in a shorter orbital period, a larger Kerr parameter and more available rotational energy.

This situation does not, however, necessarily imply a more energetic explosion or a brighter gamma-ray burst during the formation of the black hole in a system with a low-mass companion star.
In fact, a "Goldilocks" scenario may be at play.
In such a scenario, a system with a massive companion will not evolve into a very short-orbital period binary and so it will not develop a rapidly spinning black hole nor an accretion disk, thus all or most of the massive primary will collapse into the black hole (remember that as we are using the Collapsar model we assume that the initial supernova is not successful).
On the other side of the companion-mass spectrum, a system with a rather small companion (but still massive enough to remove the envelope and spiral in), may produce an extreme Kerr black hole.
Thus, with an extremely large amount of rotational energy the Kerr black hole may blow away the accretion disk too quickly to launch the gamma-ray burst.
Such black hole should be observed with a Kerr parameter close to what it was born with given that it will not tap into its rotational-energy reserves for too long as the central Blandford-Znajek engine will be rapidly blown apart by an energetic hypernova.
Hence, only a black hole born with the right heritage of rotational energy will be able to tap into its energy resources such as to launch a long gamma-ray burst and deplete a substantial amount of the Kerr parameter from the black hole.

As recent measurements of Kerr parameters \citep{2006ApJ...636L.113S,2011MNRAS.tmp.1036S} confirm, our results in at least three of our more trustworthy studied systems (GRO J1655$-$40, 4U 1543$-$47, XTE J1550$-$564 and maybe even LMC X$-$3), show that black holes with large rotational energies may easily blow away their accretion disks, and thus, prevent the central engine from triggering a gamma-ray burst.
It may also be that the gamma-ray burst is not fully prevented but it becomes, instead, subluminous.
Such situation may be similar to that of GRB 980425/SN 1998bw where the gamma-ray burst was subluminous but the hypernova was extremely energetic.
Further evidence of this may be that, as \citet{1999Natur.401..142I} have found, the companion of Nova Sco seems to have been polluted by material from the hypernova explosion at the time of black hole formation but still, the black hole seems to have little of its original spin.

In \citet{2008ApJ...685.1063B} we make the case that a companion with mass near $5 \msun$ may be on the right ballpark in order to produce a Cosmological (i.e., a bright) gamma-ray burst.
Thus, LMC X$-$3 may be the best local candidate of a black-hole binary which is a remnant of a GRB/HN explosion.
And indeed, our estimate for the Kerr parameter of this black hole is lower but not too distant from the observational estimate of \citet{2006ApJ...647..525D}.
As mentioned above, our estimate is for the initial Kerr parameter, before powering a GRB/HN explosion; it is reasonable to expect a measurement below this value if the black hole powered such an event.
Further tests that such event occurred could come from measuring the system velocity of the LMC X$-$3 binary.
This would be useful because we know the binary lost mass asymmetrically given that, even if the explosion were spherically symmetric, it occurs off the center of mass of the binary.
Hence the system received a Blaauw-Boersma kick \citep{1961BAN....15..265B,1961BAN....15..291B} and thus the system was kicked with a velocity proportional to the mass loss \citep[see, e.g., the discussion in App. A of][]{2001NewA....6..457B}.
From the energetics (obtained from the Kerr parameter), our estimate is that somewhere between 3 and $3.5 \msun$ were lost during the explosion and the system acquired a velocity of $\sim 45$ km s$^{-1}$.

To conclude with the set of seven black-hole binaries with nuclear evolved companions the difference between the observed and predicted values for the Kerr parameter in GRS 1915$+$105 must be explained.
In \citet{2011ApJ...727...29M} it is proposed that Cyg X$-$1, V4641 Sag (XTE J1819$-$254) and GRS 1915$+$105 are all binary systems at different stages of a similar evolutionary track.
That is, the companion star is more massive than the primary at the time of the collapse into the black hole (although XTE J1819$-$254 and GRS 1915$+$105 may not necessarily have ever had a companion star as massive as the one in Cyg X$-$1).
As the companion star evolves it eventually starts transferring mass to the black hole.
There can be a couple of short phases but with substantial stable-mass-transfer rates ($\sim 4\times10^{-3} \msun$ yr$^{-1}$ and $\sim 4\times10^{-4} \msun$) as modeled by \citet{2003MNRAS.341..385P}, thus allowing for hypercritical accretion of several solar masses into the black hole.
In these binaries even conservative mass transfer will expand the orbit when the mass ratio is inverted (as is the case of XTE J1819$-$254 and GRS 1915$+$105), that is, when the less massive companion transfers mass to the more massive black hole.
When the original mass of the black hole ($M_{BH} \sim 6 \msun$) is doubled, the Kerr parameter of the black hole reaches its maximum value $a_\star \sim 1$ (see the right plot of Fig.~\ref{fig:1}).
At the same time, the orbital period becomes longer, growing from a couple of days to over 30 days which is the situation presently observed in GRS 1915$+$105.
This can be easily seen from the relation of the orbital periods to the system masses for conservative mass transfer:
\be
\frac{P_{1819}}{P_{1915}} = \frac{2.8 d}{33.5 d} = \left(\frac{(M_{BH} \times m_d)_{1915}}{(M_{BH} \times m_d)_{1819}}\right)^3
= \left(\frac{14 \msun \times 1.8\msun}{6 \msun \times 9.6 \msun}\right)^3,
\ee
where $M_{BH}$ is the mass of the black hole, $m_d$ is the mass of the donor (companion) star and the subindices 1819 and 1915 represent the corresponding present orbital periods and masses for the XTE J1819$-$254 and GRS 1915$+$105 systems, respectively.
The actual mass of the companion star of GRS 1915$+$105 is actually not $1.8 \msun$ but rather $1.2 \msun$, but our restriction of conservative mass transfer may also be somewhat loosened to account for some mass loss.
It must also be noted that the natal mass of the black hole is $6 \msun$ while the companion star has $9.6 \msun$ in this example, whereas in XTE J1819$-$254 the ratio is inverted.
Nonetheless, for conservative mass transfer, during this inversion of the masses, the orbital period would initially decrease and later it would increase leaving it unchanged.
Thus, GRS 1915$+$105 makes a good case in favor of a black hole being spun up by hypercritical mass accretion.

\section{Massive BH Binaries\label{sec-MassiveBHBs}}

On the other end of the companion-mass spectrum, we have three black-hole binaries with massive companions.
In fact, in all these HMXBs (high-mass X-ray binaries) the companions are presently more massive than the black holes themselves.
These binaries are the eclipsing system M33 X$-$7, LMC X$-$1 (both of them extra-galactic) and Cyg X$-$1 (Galactic).
From the discussion above, it is expected that such systems will have wider orbits at the time of the collapse.
This is also rather intuitive, as more massive stars will have larger radii and, therefore, will only fit in binaries with larger orbital separations.
We would expect that massive binaries which end up in much shorter orbits would merge rather quickly and would not be observable for long.
Hence, one expects the binaries which survive the mergers and, therefore, are observable to have rather large orbital separations, large black hole masses (due to the lack of energy to produce an explosion after the initial supernova fails) and small values on their natal Kerr parameters.
However, it has been found that these three systems (M33 X$-$7, LMC X$-$1 and Cyg X$-$1 ) possess black holes with rather large Kerr parameters \citep[see][who measure the values of $\sim0.77$, $\sim0.87$ and $>0.97$, respectively for these sources as shown in table~\ref{tab-results}]{2008ApJ...679L..37L,2009ApJ...701.1076G,2011ApJ...742...85G}.

This is rather interesting on several aspects.  First, the necessary orbital periods for these binaries to transfer enough angular momentum to their black hole progenitors so they are born with these Kerr parameters are below $P_{orb}<0.4$ days (for $a_\star \simeq 0.8$) and $P_{orb}<0.3$ days (for $a_\star \simeq 1$; see the left plot of Fig.~\ref{fig:1}).
This implies orbital separations smaller than the stellar radii of any of the two stars in these binaries:  $a \lesssim 10 \rsun \lesssim R_{20,ZAMS}$, where $R_{20,ZAMS}$ is the ZAMS radius of a star of at least 20 $\msun$.

Second, if we want to assume the black holes acquired these Kerr parameters after they were formed, one must assume that several solar masses were transferred into the black hole in a few million years (given the short lifetime of such massive companions).
This implies that the accretion onto the black hole was at least a factor of $\sim 100$ above the Eddington limit, most likely hypercritical (hypercritical accretion occurs for $\dot{M}_{acc} \gtrsim 10^{4} \dot{M}_{Edd}$ where $\dot{M}_{Edd}$ is the Eddington-luminosity-limited accretion rate).
And, given that Roche-lobe overflow (RLOF) mass transfer would be unstable \citep[due to a mass ratio $q \equiv M_{comp}/M_{BH} > 1.28$; see ][]{1999A&A...350..928T}, the mass transfer likely occurred through a focused wind \citep[or wind Roche-lobe overflow; ][]{2007ASPC..372..397M} as suggested in \citet{2008ApJ...689L...9M} and \citet{2011MNRAS.413..183M}.  Indeed, Cyg X$-$1 has been observed \citep{1998ApJ...506..424S} to have both, radial and BH-focused wind components.

Third, if one assumes that black holes cannot be born with such a large Kerr parameter nor can they accrete such a large amount of mass and angular momentum in order to obtain it after they are formed (due to Eddington-limit restrictions and/or the instabilities in the mass transfer), then one concludes that the spin must have been acquired during the formation of the black hole.
This is the case in \citet{2010arXiv1011.4528A}.
They propose the mechanism of \citet[][]{2007Natur.445...58B} as a way to explain the large Kerr parameter in Cyg X$-$1.
However Moreno M\'endez and Cantiello (2012, in preparation) seem to find that such scenario is unlikely in massive black holes.
This is because conservation laws (for energy and angular-momentum) pose extreme requirements on the conversion efficiency of gravitational binding energy into rotational energy.

Fourth, if these Kerr parameters were natal, the available energies for the Blandford-Znajek engine would be on the order of 1,300 to over 3,000 bethes (1 B $\equiv$ 1 bethe $\equiv$ 1 foe $\equiv 10^{51}$ erg, i.e. the typical luminous energy of a supernova) for these black holes.  Most likely, such energies would be rapidly delivered to the accretion disk, quickly blowing it away, preventing the central engine from working long enough to launch a gamma-ray burst but probably producing an extremely luminous and energetic hypernova.
Most likely the companions would show metal enrichment, not unlike that of Nova Sco.
Given that these binaries should be younger than Nova Sco, they should have not had the time to {\it hide} the traces of this material by mixing it in.
It is also likely that black holes formed in such energetic fashion would lose a large amount of material early during their formation and, thus, would not grow as massive as they have.
Were this the case it is likely they would break the binary apart (if half the mass of the pre-collapse binary was lost), or, at the very least, acquire large system velocities, unlike what seems to be the case for Cyg X$-$1 according to the measurements by \citet[][]{2003Sci...300.1119M} and \citet{2011ApJ...742...83R}.

\begin{table}
\tbl{Observed masses for the compact object (CO) and the Wolf-Rayet star, orbital period and expected Kerr parameter, $a_\star$ (of the black hole formed from the collapse of the Wolf-Rayet for the four known WR-compact object binaries with present day periods), expected energy available to Blandford-Znajek, and metallicity of the binary. \newline
REFERENCES: (1) \citet{2007ApJ...669L..21P}, (2) \citet{2008ApJ...678L..17S}; (3) \citet{2010MNRAS.403L..41C}; (4) \citet{2009MNRAS.392..251H}, (5) \citet{2010ApJ...708..862K}, (6) \citet{2010ApJ...715..697K}, (7) \citet{2010ApJ...718..488S}
.}
{\begin{tabular}{|c|c|c|c|c|c|c|c|}
\hline
      WR-BH/NS       &        $M_{CO}$        &        $M_{WR}$        &  $P_{now}$ &       Model      &   $E_{BZ}$   &     Z     &   References.   \\
                     &       $[\msun]$        &        $[\msun]$       &   [days]   &   $a_{\star,2}$  &   [bethes]   & $[\Zsun]$ &                 \\
\hline\hline
\multicolumn{8}{|c|}{Extragalactic} \\
\hline
      IC~10~X$-$1    &     $32.7\pm2.6$       &        $\sim35$        &    $1.46$  &  $\gtrsim0.35$   & $\gtrsim158$ &   $0.3$   &    1, 2, 3   \\
      NGC~300~X$-$1  &         $20\pm4$       &    $26^{+7}_{-5}$      &    $1.35$  &  $\gtrsim0.4$    & $\gtrsim158$ &   $0.6$   &       3      \\
\hline
\multicolumn{8}{|c|}{Galactic} \\
\hline
      Cyg~X$-$3      &        $10 - 30$ ?     &        $3 - 57$ ?      &    $0.20$  &     $\sim1$      & $\gtrsim800$ &    $1$?    &  4, 5, 6, 7 \\
\hline
\end{tabular}
\label{Tab:WR-BH}}
\end{table}

In any case, it results interesting to know whether other black-hole binaries with massive companions have such large Kerr parameters.
So far, two more such systems have been reported, mainly differing from the above three by having Wolf-Rayet (WR) companions and smaller orbital separations.
It was found by \citet{2007ApJ...669L..21P} and later confirmed by \citet{2008ApJ...678L..17S}, that IC 10 X$-$1 has a Wolf-Rayet companion in a $P_{orb}=34.93\pm0.04$-hour orbit.
The corresponding masses estimated at $M_{BH}\ge32.7\pm2.6~\msun$ and $M_{WR}=35~\msun$ (from spectral fitting; see table~\ref{Tab:WR-BH}).
Even for a more conservative mass estimate of the Wolf-Rayet companion star of $M_{WR}=17~\msun$, the black hole appears to have a mass $M_{BH}\ge23.1\pm2.1~\msun$.
Similarly, according to \citet{2010MNRAS.403L..41C} the system formed by NGC 300 X$-$1 and WR \#41 consists of either a $20\pm 4~\msun$ black hole and a companion Wolf-Rayet star of $26^{+7}_{-5}~\msun$, or, if WR \#41 provides half of the visual continuum flux, then the masses are $15^{+4}_{-2.5}~\msun$ and $14^{+3}_{-2.5}~\msun$ respectively.
The black hole and the Wolf-Rayet orbit each other with an orbital period of $P_{orb}\!=\!32.3\pm0.2$ hours.

Even assuming conservative-mass transfer between the companion stars and the black holes of IC 10 X$-$1 and NGC 300 X$-$1, the orbital separations could not have been much smaller than what they presently are given that both systems possess such massive companions.
Thus one would expect to measure Kerr parameters for these two black holes around $a_\star \sim 0.4$ (for the lower estimates of the black hole masses).
That is, of course, unless substantial mass has been accreted by the black holes through some form or another of Roche-lobe overflow.
In which case it would not be too surprising to find them with large Kerr parameters like M33 X$-$7, LMC X$-$1 and Cyg X$-$1.

Another interesting question is whether any of these WR-BH systems may produce a GRB/HN (and a BH-BH binary) from the collapse of the massive companion star.
Were the Wolf-Rayet stars tidally locked to the black holes, and were they to collapse in the near future, they may produce gamma-ray bursts.
For the present orbital period values we could expect an $a_\star\!\sim\!0.35$ for the formation of a black hole with $M_{BH}\!=\!11~\msun$, or $a_\star\!\sim\!0.40$ for a black hole with $M_{BH}\!=\!15~\msun$.
These numbers translate into available energies between $E_{BZ}\!=\!158$ and 285 bethes.
However, if the Wolf-Rayet stars take too long to collapse, they may lose/transfer too much mass, thus increasing the orbital separation and slowing the rotation of the Wolf-Rayet (both, through mass/angular momentum loss, and through tidal synchronization).

There is another Wolf-Rayet binary which may also be a possible source from where a GRB/HN may be observed.
The Wolf-Rayet (V 1521 Cyg) in Cyg X$-$3 orbits a compact object with an orbital period of $P_{orb}=4.79$ hours.
Cygnus X$-$3 has recently gone through a series of studies which suggest that it harbors a black hole
\citep[rather than a neutron star;][]{2009MNRAS.392..251H,2010ApJ...708..862K,2010ApJ...715..697K,2010ApJ...718..488S}.
The present day orbital period translates to a Kerr parameter $a_\star\!\sim\!1$ almost regardless of the mass collapsing into a black hole, assuming such an object collapses into a black hole without losing a substantial amount of angular momentum.
The corresponding available energy would be $E_{BZ}\!=\!264(M_{BH}/\msun)$ bethes.
If we consider that a black hole with a mass of $M_{BH}\!\gtrsim\!3~\msun$ forms, this implies an energy of $E_{BZ}\!\gtrsim\!800$ bethes.
Nonetheless, with such a large amount of available rotational energy, a substantial amount of the star may be blown away before producing a gamma-ray burst or forming a black hole.
Nonetheless, the hypernova that may result from such a system may be extremely energetic.


\section{Conclusions\label{sec-Concl}}

In this proceeding the work presented in \citet{2002ApJ...575..996L,2007ApJ...671L..41B,2008ApJ...685.1063B,2008ApJ...689L...9M,2011ApJ...727...29M} and \citet{2011MNRAS.413..183M} is summarized.
In these papers, a model of stellar binary evolution is put forward where the primary star evolves as a single star until helium-shell burning begins.
This allows the formation of a massive stellar core which eventually collapses to form a black hole.
At this point, the red giant star expands an extra 10 to $15\%$ and begins a phase of unstable Case-C mass transfer (via Roche-lobe overflow) to the companion star.
The binary orbit shrinks until the companion star plunges into the envelope of the primary star and, thus, a phase of common-envelope evolution begins.
The hydrogen envelope of the primary star is ejected at the expense of losing energy from the orbit of the binary.
When the orbital separation, $a_{orb}$, has been reduced from $a_{orb} \gtrsim 1,500\rsun$ to only a few solar radii the common envelope is finally removed.
The exposed massive core of the primary star then becomes tidally synchronized with the companion star, hence it is spun up.
A strong magnetic field allows for transport of angular momentum to the core resulting in a star that rotates as a solid body.
Knowledge of the orbital period before the collapse of the primary star into a black hole allows us to produce an estimate of the available angular momentum and, therefore, the Kerr parameter and the rotational energy of the black hole.
This is the same knowledge which we use to asses whether an observed black-hole binary may be a relic of a GRB/HN explosion.

Our results seem to get observational confirmation from the measurements by \citet{2006ApJ...636L.113S} and \citet{2011MNRAS.tmp.1036S} for the Kerr parameters of GRO J1655$-$40 (Nova Sco), 4U 1543$-$47 (Il Lupi) and XTE J1550$-$564.
For these three black-hole binary systems, our predictions fall well within the margin of error of the measurements.
Our results are also consistent with the measurement of the Kerr parameter estimated by \citet{2006ApJ...647..525D} for LMC X$-$3.
Even when our predicted value is somewhat above the measurement estimate, we propose this system as a candidate of a relic of a Cosmological-type of GRB/HN explosion, the missing angular momentum from the observations being compensated by the rotational energy spent by the black hole in producing the explosive event at its birth.

Our results seem to fall short from the observations on all the HMXBs, where we predict low natal Kerr parameters and, instead, the observations show large Kerr parameters.
Nonetheless, we propose that these systems have undergone a phase of substantial mass transfer from the companion star to the black hole where the Kerr parameter of the later has been dramatically increased.
In at least one system, GRS 1915$+$105, the companion star seems to have lost most of its original mass to the black hole and, as a result, the orbital period of the binary, as well as the Kerr parameter of the black hole, have increased dramatically from the time of the collapse.
We propose that XTE J1819$-$254 may be a binary in an intermediate phase between a system similar to Cyg X$-$1 and one similar to GRS 1915$+$105.

It appears that the natal Kerr parameter of stellar mass black holes can be altered by two mechanisms.
The first one is during the gravitational collapse phase; the rotational energy of the forming black hole can be tapped through the Blandford-Znajek mechanism to produce a GRB/HN event, thus lowering the value of the Kerr parameter one observes in a black hole as seems to be the case in LMC X$-$3
(where, nonetheless, we do not expect a drastic decrease in rotation due to the fact that the energy necessary to produce even a Cosmological GRB/HN kind of event is only a small fraction of the available rotational energy).
The second mechanism involves accreting a substantial amount of mass and angular momentum into the black hole once the companion star or its wind fill their Roche lobe.

In some of the black-hole binary systems there are hints of past energetic events consistent with GRB/HN explosions.
The case of Nova Sco and its polluted companion is one such interesting example.
Close observation of the composition of black hole companions in other systems remains highly desirable.
LMC X$-$3 may well be another interesting object as it may well be the only local relic we may study from a Cosmological GRB/HN event.
Even if the distance to this black-hole binary may prevent a detailed study of the composition of the companion, further constraints on the explosion of this system may be placed through the Blaauw-Boersma kick \citep{1961BAN....15..265B,1961BAN....15..291B} as well as the fact that the binary system remained bound.

It is intriguing to consider whether other observable binaries (in particular black-hole binaries such as those of IC 10 X$-$1, NGC 300 X$-$1 or even the Galactic system Cyg X$-$3) may harbor progenitors of near-future, nearby GRBs/HNe.

\clearpage


\begin{thebibliography}{57}
\expandafter\ifx\csname natexlab\endcsname\relax\def\natexlab#1{#1}\fi
\providecommand{\enquote}[1]{``#1''}
\expandafter\ifx\csname url\endcsname\relax
  \def\url#1{\texttt{#1}}\fi
\expandafter\ifx\csname urlprefix\endcsname\relax\def\urlprefix{URL }\fi
\providecommand{\eprint}[2][]{\url{#2}}

\bibitem[{Woosley}(1993)]{1993ApJ...405..273W}
S.~E. {Woosley}, \emph{\apj} \textbf{405}, 273--277 (1993).

\bibitem[{MacFadyen} and {Woosley}(1999)]{1999ApJ...524..262M}
A.~I. {MacFadyen}, and S.~E. {Woosley}, \emph{\apj} \textbf{524}, 262--289
  (1999), \eprint{arXiv:astro-ph/9810274}.

\bibitem[{Hjorth} and {Bloom}(2011)]{2011arXiv1104.2274H}
J.~{Hjorth}, and J.~S. {Bloom}, \emph{ArXiv e-prints}  (2011),
  \eprint{1104.2274}.

\bibitem[{Paczynski}(1998)]{1998ApJ...494L..45P}
B.~{Paczynski}, \emph{\apjl} \textbf{494}, L45 (1998),
  \eprint{arXiv:astro-ph/9710086}.

\bibitem[{Blandford} and {Znajek}(1977)]{1977MNRAS.179..433B}
R.~D. {Blandford}, and R.~L. {Znajek}, \emph{\mnras} \textbf{179}, 433--456
  (1977).

\bibitem[{Bethe} and {Brown}(1998)]{1998ApJ...506..780B}
H.~A. {Bethe}, and G.~E. {Brown}, \emph{\apj} \textbf{506}, 780--789 (1998),
  \eprint{arXiv:astro-ph/9802084}.

\bibitem[{Bethe} and {Brown}(1999)]{1999ApJ...517..318B}
H.~A. {Bethe}, and G.~E. {Brown}, \emph{\apj} \textbf{517}, 318--327 (1999),
  \eprint{arXiv:astro-ph/9805355}.

\bibitem[{Lee} et~al.(2000)]{2000PhR...325...83L}
H.~K. {Lee}, R.~A.~M.~J. {Wijers}, and G.~E. {Brown}, \emph{\physrep}  (2000),
  \eprint{arXiv:astro-ph/9906213}.

\bibitem[{Lee} et~al.(2002)]{2002ApJ...575..996L}
C.~{Lee}, G.~E. {Brown}, and R.~A.~M.~J. {Wijers}, \emph{\apj} \textbf{575},
  996--1006 (2002), \eprint{arXiv:astro-ph/0109538}.

\bibitem[{Braun} and {Langer}(1995)]{1995A&A...297..483B}
H.~{Braun}, and N.~{Langer}, \emph{\aap} \textbf{297}, 483 (1995).

\bibitem[{Brown} et~al.(2007)]{2007ApJ...671L..41B}
G.~E. {Brown}, C.~{Lee}, and E.~{Moreno M{\'e}ndez}, \emph{\apjl} \textbf{671},
  L41--L44 (2007), \eprint{0707.4008}.

\bibitem[{de Mink} et~al.(2009)]{2009A&A...497..243D}
S.~E. {de Mink}, M.~{Cantiello}, N.~{Langer}, O.~R. {Pols}, I.~{Brott}, and
  S.~{Yoon}, \emph{A\&A} \textbf{497}, 243--253 (2009), \eprint{0902.1751}.

\bibitem[{Heger} et~al.(2005)]{2005ApJ...626..350H}
A.~{Heger}, S.~E. {Woosley}, and H.~C. {Spruit}, \emph{\apj} \textbf{626},
  350--363 (2005), \eprint{arXiv:astro-ph/0409422}.

\bibitem[{Yoon} et~al.(2006)]{2006A&A...460..199Y}
S.-C. {Yoon}, N.~{Langer}, and C.~{Norman}, \emph{\aap} \textbf{460}, 199--208
  (2006), \eprint{arXiv:astro-ph/0606637}.

\bibitem[{Spruit}(2002)]{2002A&A...381..923S}
H.~C. {Spruit}, \emph{A\&A} \textbf{381}, 923--932 (2002),
  \eprint{arXiv:astro-ph/0108207}.

\bibitem[{Brown} et~al.(2008)]{2008ApJ...685.1063B}
G.~E. {Brown}, C.~{Lee}, and E.~{Moreno M{\'e}ndez}, \emph{\apj} \textbf{685},
  1063--1068 (2008), \eprint{0801.0477}.

\bibitem[{Moreno M{\'e}ndez} et~al.(2008)]{2008ApJ...689L...9M}
E.~{Moreno M{\'e}ndez}, G.~E. {Brown}, C.~{Lee}, and I.~H. {Park}, \emph{\apjl}
  \textbf{689}, L9--L12 (2008), \eprint{0809.2146}.

\bibitem[{Moreno M{\'e}ndez} et~al.(2011)]{2011ApJ...727...29M}
E.~{Moreno M{\'e}ndez}, G.~E. {Brown}, C.~{Lee}, and F.~M. {Walter},
  \emph{\apj} \textbf{727}, 29 (2011).

\bibitem[{Moreno M{\'e}ndez}(2011)]{2011MNRAS.413..183M}
E.~{Moreno M{\'e}ndez}, \emph{\mnras} \textbf{413}, 183--189 (2011),
  \eprint{1011.6385}.

\bibitem[{Zahn}(1975)]{1975A&A....41..329Z}
J.-P. {Zahn}, \emph{\aap} \textbf{41}, 329--344 (1975).

\bibitem[{Zahn}(1977)]{1977A&A....57..383Z}
J.-P. {Zahn}, \emph{\aap} \textbf{57}, 383--394 (1977).

\bibitem[{van den Heuvel} and {Yoon}(2007)]{2007Ap&SS.311..177V}
E.~P.~J. {van den Heuvel}, and S.-C. {Yoon}, \emph{\apss} \textbf{311},
  177--183 (2007), \eprint{0704.0659}.

\bibitem[{Woosley} and {heger}(2011)]{2011arXiv1110.3842W}
S.~E. {Woosley}, and A.~{heger}, \emph{ArXiv e-prints}  (2011),
  \eprint{1110.3842}.

\bibitem[{Brown} et~al.(2000)]{2000NewA....5..191B}
G.~E. {Brown}, C.-H. {Lee}, R.~A.~M.~J. {Wijers}, H.~K. {Lee}, G.~{Israelian},
  and H.~A. {Bethe}, \emph{\na} \textbf{5}, 191--210 (2000),
  \eprint{arXiv:astro-ph/0003361}.

\bibitem[{Corral-Santana} et~al.(2011)]{2011MNRAS.413L..15C}
J.~M. {Corral-Santana}, J.~{Casares}, T.~{Shahbaz}, C.~{Zurita}, I.~G.
  {Mart{\'{\i}}nez-Pais}, and P.~{Rodr{\'{\i}}guez-Gil}, \emph{\mnras}
  \textbf{413}, L15--L19 (2011), \eprint{1102.0654}.

\bibitem[{Gou} et~al.(2010)]{2010ApJ...718L.122G}
L.~{Gou}, J.~E. {McClintock}, J.~F. {Steiner}, R.~{Narayan}, A.~G. {Cantrell},
  C.~D. {Bailyn}, and J.~A. {Orosz}, \emph{\apjl} \textbf{718}, L122--L126
  (2010), \eprint{1002.2211}.

\bibitem[{Steiner} et~al.(2011)]{2011MNRAS.tmp.1036S}
J.~F. {Steiner}, R.~C. {Reis}, J.~E. {McClintock}, R.~{Narayan}, R.~A.
  {Remillard}, J.~A. {Orosz}, L.~{Gou}, A.~C. {Fabian}, and M.~A.~P. {Torres},
  \emph{\mnras} p. 1036 (2011), \eprint{1010.1013}.

\bibitem[{Reid} et~al.(2011)]{2011ApJ...742...83R}
M.~J. {Reid}, J.~E. {McClintock}, R.~{Narayan}, L.~{Gou}, R.~A. {Remillard},
  and J.~A. {Orosz}, \emph{\apj} \textbf{742}, 83 (2011), \eprint{1106.3688}.

\bibitem[{Orosz} et~al.(2011)]{2011ApJ...742...84O}
J.~A. {Orosz}, J.~E. {McClintock}, J.~P. {Aufdenberg}, R.~A. {Remillard}, M.~J.
  {Reid}, R.~{Narayan}, and L.~{Gou}, \emph{\apj} \textbf{742}, 84 (2011),
  \eprint{1106.3689}.

\bibitem[{Gou} et~al.(2011)]{2011ApJ...742...85G}
L.~{Gou}, J.~E. {McClintock}, M.~J. {Reid}, J.~A. {Orosz}, J.~F. {Steiner},
  R.~{Narayan}, J.~{Xiang}, R.~A. {Remillard}, K.~A. {Arnaud}, and S.~W.
  {Davis}, \emph{\apj} \textbf{742}, 85 (2011), \eprint{1106.3690}.

\bibitem[{Orosz} et~al.(2009)]{2009ApJ...697..573O}
J.~A. {Orosz}, D.~{Steeghs}, J.~E. {McClintock}, M.~A.~P. {Torres},
  I.~{Bochkov}, L.~{Gou}, R.~{Narayan}, M.~{Blaschak}, A.~M. {Levine}, R.~A.
  {Remillard}, C.~D. {Bailyn}, M.~M. {Dwyer}, and M.~{Buxton}, \emph{\apj}
  \textbf{697}, 573--591 (2009), \eprint{0810.3447}.

\bibitem[{Gou} et~al.(2009)]{2009ApJ...701.1076G}
L.~{Gou}, J.~E. {McClintock}, J.~{Liu}, R.~{Narayan}, J.~F. {Steiner}, R.~A.
  {Remillard}, J.~A. {Orosz}, S.~W. {Davis}, K.~{Ebisawa}, and E.~M.
  {Schlegel}, \emph{\apj} \textbf{701}, 1076--1090 (2009), \eprint{0901.0920}.

\bibitem[{Orosz}(2003)]{2003IAUS..212..365O}
J.~A. {Orosz}, \enquote{{Inventory of black hole binaries},} in \emph{A Massive
  Star Odyssey: From Main Sequence to Supernova}, edited by {K.~van der Hucht,
  A.~Herrero, \& C.~Esteban}, 2003, vol. 212 of \emph{IAU Symposium}, p. 365,
  \eprint{arXiv:astro-ph/0209041}.

\bibitem[{Davis} et~al.(2006)]{2006ApJ...647..525D}
S.~W. {Davis}, C.~{Done}, and O.~M. {Blaes}, \emph{\apj} \textbf{647}, 525--538
  (2006), \eprint{arXiv:astro-ph/0602245}.

\bibitem[{Pietsch} et~al.(2006)]{2006ApJ...646..420P}
W.~{Pietsch}, F.~{Haberl}, M.~{Sasaki}, T.~J. {Gaetz}, P.~P. {Plucinsky},
  P.~{Ghavamian}, K.~S. {Long}, and T.~G. {Pannuti}, \emph{\apj} \textbf{646},
  420--428 (2006), \eprint{arXiv:astro-ph/0603698}.

\bibitem[{Orosz} et~al.(2007)]{2007Natur.449..872O}
J.~A. {Orosz}, J.~E. {McClintock}, R.~{Narayan}, C.~D. {Bailyn}, J.~D.
  {Hartman}, L.~{Macri}, J.~{Liu}, W.~{Pietsch}, R.~A. {Remillard},
  A.~{Shporer}, and T.~{Mazeh}, \emph{\nat} \textbf{449}, 872--875 (2007),
  \eprint{0710.3165}.

\bibitem[{Liu} et~al.(2008)]{2008ApJ...679L..37L}
J.~{Liu}, J.~E. {McClintock}, R.~{Narayan}, S.~W. {Davis}, and J.~A. {Orosz},
  \emph{\apjl} \textbf{679}, L37--L40 (2008), \eprint{0803.1834}.

\bibitem[{Moreno Mendez} et~al.(2006)]{2006astro.ph.12461M}
E.~{Moreno Mendez}, G.~E. {Brown}, C.~. {Lee}, and F.~M. {Walter}, \emph{ArXiv
  Astrophysics e-prints}  (2006), \eprint{arXiv:astro-ph/0612461}.

\bibitem[{Shafee} et~al.(2006)]{2006ApJ...636L.113S}
R.~{Shafee}, J.~E. {McClintock}, R.~{Narayan}, S.~W. {Davis}, L.~{Li}, and
  R.~A. {Remillard}, \emph{\apjl} \textbf{636}, L113--L116 (2006),
  \eprint{arXiv:astro-ph/0508302}.

\bibitem[{Israelian} et~al.(1999)]{1999Natur.401..142I}
G.~{Israelian}, R.~{Rebolo}, G.~{Basri}, J.~{Casares}, and E.~L.
  {Mart{\'{\i}}n}, \emph{\nat} \textbf{401}, 142--144 (1999).

\bibitem[{Blaauw}(1961)]{1961BAN....15..265B}
A.~{Blaauw}, \emph{\bain} \textbf{15}, 265 (1961).

\bibitem[{Boersma}(1961)]{1961BAN....15..291B}
J.~{Boersma}, \emph{\bain} \textbf{15}, 291--301 (1961).

\bibitem[{Brown} et~al.(2001)]{2001NewA....6..457B}
G.~E. {Brown}, A.~{Heger}, N.~{Langer}, C.-H. {Lee}, S.~{Wellstein}, and H.~A.
  {Bethe}, \emph{\na} \textbf{6}, 457--470 (2001),
  \eprint{arXiv:astro-ph/0102379}.

\bibitem[{Podsiadlowski} et~al.(2003)]{2003MNRAS.341..385P}
P.~{Podsiadlowski}, S.~{Rappaport}, and Z.~{Han}, \emph{\mnras} \textbf{341},
  385--404 (2003), \eprint{arXiv:astro-ph/0207153}.

\bibitem[{Tauris} and {Savonije}(1999)]{1999A&A...350..928T}
T.~M. {Tauris}, and G.~J. {Savonije}, \emph{\aap} \textbf{350}, 928--944
  (1999), \eprint{arXiv:astro-ph/9909147}.

\bibitem[{Mohamed} and {Podsiadlowski}(2007)]{2007ASPC..372..397M}
S.~{Mohamed}, and P.~{Podsiadlowski}, \enquote{{Wind Roche-Lobe Overflow: a New
  Mass-Transfer Mode for Wide Binaries},} in \emph{15th European Workshop on
  White Dwarfs}, edited by {R.~Napiwotzki \& M.~R.~Burleigh}, 2007, vol. 372 of
  \emph{Astronomical Society of the Pacific Conference Series}, p. 397.

\bibitem[{Sowers} et~al.(1998)]{1998ApJ...506..424S}
J.~W. {Sowers}, D.~R. {Gies}, W.~G. {Bagnuolo}, A.~W. {Shafter}, R.~{Wiemker},
  and M.~S. {Wiggs}, \emph{\apj} \textbf{506}, 424--430 (1998).

\bibitem[{Axelsson} et~al.(2010)]{2010arXiv1011.4528A}
M.~{Axelsson}, R.~P. {Church}, M.~B. {Davies}, A.~J. {Levan}, and F.~{Ryde},
  \emph{ArXiv e-prints}  (2010), \eprint{1011.4528}.

\bibitem[{Blondin} and {Mezzacappa}(2007)]{2007Natur.445...58B}
J.~M. {Blondin}, and A.~{Mezzacappa}, \emph{\nat} \textbf{445}, 58--60 (2007),
  \eprint{arXiv:astro-ph/0611680}.

\bibitem[{Mirabel} and {Rodrigues}(2003)]{2003Sci...300.1119M}
I.~F. {Mirabel}, and I.~{Rodrigues}, \emph{Science} \textbf{300}, 1119--1121
  (2003), \eprint{arXiv:astro-ph/0305205}.

\bibitem[{Prestwich} et~al.(2007)]{2007ApJ...669L..21P}
A.~H. {Prestwich}, R.~{Kilgard}, P.~A. {Crowther}, S.~{Carpano}, A.~M.~T.
  {Pollock}, A.~{Zezas}, S.~H. {Saar}, T.~P. {Roberts}, and M.~J. {Ward},
  \emph{\apjl} \textbf{669}, L21--L24 (2007), \eprint{0709.2892}.

\bibitem[{Silverman} and {Filippenko}(2008)]{2008ApJ...678L..17S}
J.~M. {Silverman}, and A.~V. {Filippenko}, \emph{\apjl} \textbf{678}, L17--L20
  (2008), \eprint{0802.2716}.

\bibitem[{Crowther} et~al.(2010)]{2010MNRAS.403L..41C}
P.~A. {Crowther}, R.~{Barnard}, S.~{Carpano}, J.~S. {Clark}, V.~S. {Dhillon},
  and A.~M.~T. {Pollock}, \emph{\mnras} \textbf{403}, L41--L45 (2010),
  \eprint{1001.4616}.

\bibitem[{Hjalmarsdotter} et~al.(2009)]{2009MNRAS.392..251H}
L.~{Hjalmarsdotter}, A.~A. {Zdziarski}, A.~{Szostek}, and D.~C. {Hannikainen},
  \emph{\mnras} \textbf{392}, 251--263 (2009), \eprint{0810.1685}.

\bibitem[{Karak} et~al.(2010{\natexlab{a}})]{2010ApJ...708..862K}
B.~B. {Karak}, J.~{Dutta}, and B.~{Mukhopadhyay}, \emph{\apj} \textbf{708},
  862--867 (2010{\natexlab{a}}), \eprint{0911.1701}.

\bibitem[{Karak} et~al.(2010{\natexlab{b}})]{2010ApJ...715..697K}
B.~B. {Karak}, J.~{Dutta}, and B.~{Mukhopadhyay}, \emph{\apj} \textbf{715},
  697--+ (2010{\natexlab{b}}).

\bibitem[{Shrader} et~al.(2010)]{2010ApJ...718..488S}
C.~R. {Shrader}, L.~{Titarchuk}, and N.~{Shaposhnikov}, \emph{\apj}
  \textbf{718}, 488--493 (2010), \eprint{1005.5362}.

\end{thebibliography}
\end{document}